\documentclass{article}

\usepackage{bm}
\usepackage[psamsfonts]{amsfonts}
\usepackage{amsthm}
\usepackage{amssymb}
\usepackage{amsmath}
\usepackage{amsopn}
\usepackage{graphicx}
\usepackage{bm}
\usepackage{caption}

\theoremstyle{plain}

\DeclareMathOperator{\Complex}{\mathbb{C}}
\DeclareMathOperator{\Real}{\mathbb{R}}

 \DeclareMathOperator{\Mat}{Mat}
 
\DeclareMathOperator{\Tr}{Tr}

\allowdisplaybreaks

\title{Topological excitations  in 2D spin system with high spin $s\geqslant 1$}
\author{\Large Julia Bernatska, Petro Holod}
\date{}
\begin{document}

\maketitle

\begin{abstract}
We construct a class of topological excitations of a mean field in a
two-dimensional spin system represented by a quantum Heisenberg
model with high powers of exchange interaction. The quantum model is
associated with a classical one (the continuous classical analogue)
that is based on a Landau-Lifshitz like equation, and describes
large-scale fluctuations of the mean field. On the other hand, the
classical model is a Hamiltonian system on a coadjoint orbit of the
unitary group SU($2s\,{+}\,1$) in the case of spin~$s$. We have
found a class of mean field configurations that can be interpreted
as topological excitations, because they have fixed topological
charges. Such excitations change their shapes and grow preserving an
energy.

\end{abstract}

\section{Introduction}
According to Mermin and Wagner~\cite{Mermin} there is no
ferromagnetic or antiferromagnetic order in the one- and
two-dimensional isotropic Heisenberg models with interactions of
finite range at nonzero temperature. This statement is proven due to
Bogoliubov's inequality in the general case. Here we construct
excitations that cause a destruction of a long-range nematic or
mixed ferromagnetic-nematic order. This is an extension of the
results of Belavin and Polyakov \cite{Belavin}.

We model a planar magnet by a square atomic lattice with the same
spin~$s$ at each site. We describe this two-dimensional spin system
by a generalized Heisenberg Hamiltonian, taking into account high
powers of the exchange interaction
$(\hat{\bm{S}}_n,\hat{\bm{S}}_{n+\delta})$, where $\hat{\bm{S}}_n$
is a vector of spin operators at site $n$. By a mean field
approximation we obtain a classical long-range equation from the
quantum Heisenberg one.

An equation for a mean field (the field of magnetization and
multipole moments) is a Hamiltonian equation on a coadjoint orbit of
Lie group. At the same time, this is a generalization of the
well-known Landau-Lifshitz equation for a magnetization field. In
this context we obtain effective Hamiltonians for the magnetic
system in question. Using  K\"{a}hlerian structure of coadjoint
orbits, we construct effective Hamiltonians such that their minimums
are proportional to topological charges of excitations. In addition,
we produce these mean field configurations that give minimums to the
Hamiltonians.

%------------------------------------------------------------------------------
\section{Quantum and classical models}
As mentioned above, we represent the spin system by a planar atomic
lattice with the same spin $s$ at all sites. We assign three spin
operators $(\hat{S}_n^1,\,\hat{S}_n^2,\,\hat{S}_n^3)=\hat{\bm{S}}_n$
to each atom $n$; the operators obey the standard commutation
relations
\begin{equation*}
[\hat{S}^a_n,\,\hat{S}^b_m] = i\varepsilon_{abc} \hat{S}^c_n
\delta_{nm},
\end{equation*}
where $a,$ $b,$ $c$ run over the values $\{ 1,\,2,\,3\}$, and
$\delta_{nm}$ is the Kronecker symbol.

%------------------------------------------------------------------------------
\subsection{Generalized Heisenberg Hamiltonians}
We are interested in so called high spins $s\,{\geqslant}\, 1$. In
this case, we can describe the system by the following
\emph{bilinear-biquadratic Hamiltonian}
\begin{equation*}
\hat{\mathcal{H}}^{2} = -\sum_{n,\delta} \Bigl(J
(\hat{\bm{S}}_n,\hat{\bm{S}}_{n+\delta}) + K
(\hat{\bm{S}}_n,\hat{\bm{S}}_{n+\delta})^2 \Bigr).
\end{equation*}
Here $\delta$ runs over the nearest-neighbour sites, $n$ runs over
all sites of the lattice, constants $J$ and $K$ denote exchange
integrals. In the case of spin $s\,{\geqslant}\,3/2$, the above
Hamiltonian can include also the bicubic exchange, namely
\begin{equation*}
\hat{\mathcal{H}}^{3} = -\sum_{n,\delta} \Bigl(J
(\hat{\bm{S}}_n,\hat{\bm{S}}_{n+\delta}) + K
(\hat{\bm{S}}_n,\hat{\bm{S}}_{n+\delta})^2 + L
(\hat{\bm{S}}_n,\hat{\bm{S}}_{n+\delta})^3\Bigr),
\end{equation*}
where $L$ denotes the corresponding exchange integral.

One can easily write a generalized Heisenberg Hamiltonian for the
system of an arbitrary spin $s$ or greater than $s$. This
Hamiltonian contains all powers of the exchange interaction up to
$2s$. It can be reduced to a bilinear form if one takes the
$2s\,{+}\,1$-dimensional space of irreducible representation of the
group SU(2).  The spin operators $\{\hat{S}_n^a\}$ over this space
generate  a complete associative matrix algebra, which has a
sufficient number of operators to reduce the corresponding
Hamiltonian to a bilinear form.

For example, in the case of spin $s\,{=}\,1$ the appropriate
 space of representation is 3-di\-men\-sio\-nal, and we choose a canonical basis in
the form: $\{|{+}1\rangle,\, |{-}1\rangle,\, |0\rangle\}$. The spin
operators $\{\hat{S}_n^a\}$ generate the algebra $\Mat_{3\times 3}$.
In order to form a basis in the algebra we take the tensor operators
of weight~2
\begin{equation}\label{QuadOp2}
\begin{split}
&\hat{Q}^{ab}_n = \hat{S}^a_n \hat{S}^b_n + \hat{S}^b_n
\hat{S}^a_n,\quad a\neq b,\\
&\hat{Q}^{22}_n = (\hat{S}^1_n)^2 - (\hat{S}^2_n)^2, \qquad
\hat{Q}^{20}_n = \sqrt{3}\bigl((\hat{S}^3_n)^2 - \tfrac{2}{3}\bigr)
\end{split}
\end{equation}
in addition to the spin operators. The introduced operators are
called \emph{quadrupole operators}.

In the case of spin $s\,{=}\,3/2$ the appropriate space of
representation is 4{-}dimensional, and $\{|{+}\tfrac{3}{2}\rangle,
|{+}\tfrac{1}{2}\rangle, |{-}\tfrac{1}{2}\rangle,
|{-}\tfrac{3}{2}\rangle\}$ is a canonical basis. We complete the
associative matrix algebra $\Mat_{4\times 4}$ of $\{\hat{S}_n^a\}$
by the tensor operators of weights 2 and 3, defining them by the
following formulas:
\begin{equation}\label{QuadOp3}
\begin{split}
&\hat{Q}^{ab}_n = \tfrac{\sqrt{5}}{2\sqrt{3}}\Bigl(\hat{S}^a_n
\hat{S}^b_n + \hat{S}^b_n \hat{S}^a_n\Bigr),\quad a\neq b\\
&\hat{Q}^{22}_n =  \tfrac{\sqrt{5}}{2\sqrt{3}} \Bigl((\hat{S}^1_n)^2
- (\hat{S}^2_n)^2\Bigr), \quad \hat{Q}^{20}_n =
\tfrac{\sqrt{5}}{2}\bigl((\hat{S}^3_n)^2 - \tfrac{5}{4}\bigr),
\end{split}
\end{equation}
\begin{equation}\label{SextOp3}
\begin{split}
&\hat{T}^{a3}_n = (\hat{Q}^{a2}_n \hat{S}^3_n +\hat{S}^3_n
\hat{Q}^{a2}_n ), \quad a,b \in \{1,2\},\ a\neq b,\\
&\hat{T}^{ab}_n =
\tfrac{1}{\sqrt{6}}\Bigl((\hat{S}^a_n)^2\hat{S}^b_n
+\hat{S}^b_n(\hat{S}^a_n)^2 + \hat{S}^a_n\hat{S}^b_n\hat{S}^a_n -
(\hat{S}^b_n)^3 \Bigr),\\
&\hat{T}^{3a}_n = \tfrac{1}{\sqrt{10}}\Bigl( \hat{Q}^{a3}_n
\hat{S}^3_n + \hat{S}^3_n \hat{Q}^{a3}_n +
\sqrt{3}(\hat{Q}^{20}_n\hat{S}^a_n + \hat{S}^a_n
\hat{Q}^{20}_n) \Bigr), \\
&\hat{T}^{30}_n = \tfrac{1}{12}\bigl(41\hat{S}^3_n -
20(\hat{S}^3_n)^3\bigr).
\end{split}
\end{equation}
We call the tensor operators  of weight 3 \emph{sextupole
operators}. In what follows we denote all tensor operators over the
chosen space of representation by $\{\hat{P}^a_n\}$.

In terms of representation operators a generalized Heisenberg
Hamiltonian gets a bilinear form. For the Hamiltonians considered
above we have:
\begin{align*}
&\hat{\mathcal{H}}^{2} = -(J{-}\tfrac{1}{2}K)\sum_{n,\delta}
\sum_{b} \hat{S}_n^{b}
\hat{S}_{n+\delta}^{b}-\tfrac{1}{2}K\sum_{n,\delta} \sum_{\alpha}
\hat{Q}_n^{\alpha} \hat{Q}_{n+\delta}^{\alpha} - \tfrac{4}{3}KN;\\
&\hat{\mathcal{H}}^{3} = -(J{-}\tfrac{1}{2}K {+} \tfrac{587}{80}
L)\sum_{n,\delta} \sum_{b} \hat{S}_n^{b} \hat{S}_{n+\delta}^{b} -
\tfrac{75}{32}(4K{-}L)N -
\\ &\phantom{\hat{\mathcal{H}}^{\text{spin 3/2}} = -}
-\tfrac{6}{5}(K{-}2L)\sum_{n,\delta} \sum_{\alpha}
\hat{Q}_n^{\alpha} \hat{Q}_{n+\delta}^{\alpha} -
\tfrac{9}{10}L\sum_{n,\delta} \sum_{\beta} \hat{T}_n^{\beta}
\hat{T}_{n+\delta}^{\beta},
\end{align*}
where $N$ is the overall number of sites in the lattice. Obviously,
the obtained bilinear Hamiltonians are SU(2)-invariant, because they
are constructed from representation operators of the group SU(2).

%------------------------------------------------------------------------------
\subsection{Mean field approximation}\label{s:MFA}
Here a mean field is a field of expectation values for the operators
$\{\hat{P}_n^{a}\}$ calculated after spontaneous breaking of
symmetry.  The breaking of symmetry is performed by switching on an
external magnetic field, which specifies an order in the system;
then the external field vanishes. Such kind of averages is also
called \emph{quasiaverages} \cite{Bogolyubov}.

We denote a mean field averaging by $\langle\cdot \rangle$, and
components of the mean field by $\{\mu_a
(\bm{x}_n)\,{=}\,\langle\hat{P}^{a}_n\rangle\}$. A mean field
approximation of the bilinear-biquadratic Hamiltonian has the form
\begin{equation*}
\hat{\mathcal{H}}^{2}_{\text{MF}} \,{=}\, {-}(J{-}\tfrac{1}{2}K)z
\sum_{n}\sum_{a=1}^3 \hat{P}_n^{a} \mu_a(\bm{x}_n)- \tfrac{1}{2}Kz
\sum_{n}\sum_{a=4}^8 \hat{P}_n^{a} \mu_a(\bm{x}_n) -
\tfrac{4}{3}KzN,
\end{equation*}
where $z$ is a number of the nearest-neighbour sites.

Evidently, a mean field Hamiltonian remains
$\mathrm{SU(2)}$-invariant. Then by an action of the group
$\mathrm{SU(2)}$ it can be reduced to a diagonal form. Of course,
this reduction is possible only in the case of thermodynamical
equilibrium and an infinite lattice, when the mean field becomes
constant and the dependance on site $n$ can be omitted. Moreover,
almost all components of the mean field vanish, except the
components corresponding to diagonal operators of $\{\hat{P}^a_n\}$.
For the bilinear-biquadratic Hamiltonian a diagonal form is the
following:
\begin{align*}
&\hat{\mathcal{H}}^{2}_{\text{MF}} \,{=}\, {-}zN
\left((J{-}\tfrac{1}{2}K) \hat{S}^{3} \mu_3+\tfrac{1}{2}K
\hat{Q}^{20} \mu_8 + \tfrac{4}{3}K \right).
\end{align*}
The remaining components are suitable to serve as \emph{order
parameters}.  The component $\mu_3$ describes a normalized
\emph{magnetization} (a ratio of the $z$-projection of magnetic
moment to a saturation magnetization). The components  $\mu_8$ and
$\mu_{15}$ are normalized \emph{projections of quadrupole} and
\emph{sextupole moments}, respectively.

Possible values of order parameters can be obtained from the
\emph{self-consistent equations}
$$\mu_a \,{=}\, \langle\hat{P}^a\rangle_{\text{MF}}
\,{=}\, \frac{\Tr \hat{P}^a e^{-\frac{\hat{h}_{\text{MF}}}{kT}}}{\Tr
e^{-\frac{\hat{h}_{\text{MF}}}{kT}}}$$ for all diagonal operators
$\hat{P}^a$. Here we use the density matrix with the one-site
Hamiltonian $\hat{h}_{\text{MF}} \,{=}\,
\hat{\mathcal{H}}_{\text{MF}}/N$. Note, that for the standard
Heisenberg Hamiltonian, when only the spin operators are considered,
a self-consistent equation turns into the well-known Weiss equation.

In the case of bilinear-biquadratic Hamiltonian, an analysis of
self-consistent equations gives the following. We adduce probable
values of order parameters in the limit $T\,{\to}\, 0$ as $J,\,
K\,{>}\,0$: 1)~$|\mu_3|\,{=}\,1,\ \mu_8 \,{=}\,
\frac{2J-K}{\sqrt{3}\, K}$; 2)~$|\mu_3|\,{=}\,\frac{1}{2},\ \mu_8
\,{=}\, \frac{J-K/2}{\sqrt{3}\, K}$; 3)~$\mu_3\,{=}\,0,\ |\mu_8|
\,{=}\, \frac{2}{\sqrt{3}}$; 4)~$\mu_3\,{=}\,0$, $|\mu_8| \,{=}\,
\frac{1}{\sqrt{3}}$. Two of the solutions correspond to states with
the ferromagnetic order, because they have a nonzero
magnetization~$\mu_3$. The other two solutions correspond to states
with the quadrupole order (so called nematic states), which are
states with zero magnetization. Solutions  2) and 4) are unstable
and called partly ordered. If $K$ becomes negative, only solution 1)
remains. These results accord with the results of \cite{Matveev,
Nauciel} and with the phase diagram of ordered states in a
one-dimensional spin system from \cite{Buchta}.

In what follows we consider an SU(3)-invariant system with the
bilinear-bi\-quad\-ra\-tic Hamiltonian. The SU(3)-invariance is
reached by assigning $J\,{=}\,K$; this is the boundary line between
the ferromagnetic and the nematic regions (see~\cite{Buchta}). It
means the system can appear in the both states. In the case of
Hamiltonian $\hat{\mathcal{H}}^3$, the maximal SU(4)-invariance is
reached as $J\,{=}\,{-}\frac{81}{44}\,K \,{=}\,
{-}\frac{81}{16}\,L$, that is located within the ferromagnetic
region.

Now we apply the mean field averaging to the quantum Heisenberg
equation
\begin{equation}\label{HeisEq}
i\hbar \frac{d \hat{P}^a_n}{dt} = [\hat{P}^a_n,\hat{\mathcal{H}}].
\end{equation}
The averaging is performed with the assumption of zero correlations
between fluctuations of $\{\hat{P}^a_n\}$ at distinct sites:
$\langle \hat{P}^a_n \hat{P}^b_m\rangle \,{=}\, \langle \hat{P}^a_n
\rangle \langle \hat{P}^b_m \rangle$. Then we take a large-scale
limit and obtain the Landau-Lifshitz like equation
\begin{equation}\label{LLEq}
\hbar \frac{\partial \mu_a}{\partial t} = 2Jl^2\, C_{abc} \mu_b
(\mu_{c,xx} + \mu_{c,yy}),
\end{equation}
where $C_{abc}$ are structure constants of the Lie algebra of
 $\{\hat{P}^a_n\}$ with the commutation relations
$[\hat{P}^a_n, \hat{P}^b_m]=iC_{abc}\hat{P}^c_n \delta_{nm}$.
Equation \eqref{LLEq} is an equation of motion for the mean field
$\{\mu_a(\bm{x})\}$ over the plain $\{\bm{x}=(x,y)\mid x,\,
y\,{\in}\,\Real\}$ that replaces the lattice.

In the case of standard Heisenberg Hamiltonian \eqref{LLEq}
coincides with the Landau-Lifshitz equation for an isotropic magnet.
Therefore, in the general case we call \eqref{LLEq} a
\emph{generalized Landau-Lifshitz equation} for the vector field
$\{\mu_a\}$. The vector field has 8 components if one exploits the
bilinear-biquadratic Hamiltonian, and 15 components for the
Hamiltonian with the bicubic exchange.

%------------------------------------------------------------------------------
\subsection{Effective Hamiltonians on coadjoint orbits}\label{s:EffAM}

The generalized Landau-Lifshitz equation \eqref{LLEq} can be
interpreted as a \emph{Hamiltonian equation on a coadjoint orbit of
Lie group}. In the case of spin  $s\,{=}\, 1$ we deal with the group
$\mathrm{SU}(3)$, in the case of an arbitrary spin $s$ the group is
 $\mathrm{SU}(2s{+}1)$. Note, that the matrices
$\{\hat{P}^a\}$ serve as a basis in the corresponding Lie algebra
$\mathfrak{su}(2s{+}1)$, and components of the mean field
$\{\mu_a\}$ serve as coordinates in the dual space to
$\mathfrak{su}(2s{+}1)$.

We start with brief description of the groups SU(3) and SU(4). For
more material see, in particular, \cite{Picken, Bernatska}.

The group SU(3) has two types of orbits: the generic
$\frac{\mathrm{SU}(3)}{\mathrm{U}(1)\times \mathrm{U}(1)}$ of
dimension~6, and the degenerate
$\frac{\mathrm{SU}(3)}{\mathrm{SU}(2)\times \mathrm{U}(1)}$ of
dimension 4. Each orbit of SU(3) is defined by two numbers $m$ and
$q$, which are values of the coordinates $\mu_3$ and $\mu_8$ at an
initial point (a point in the positive Weyl chamber).
Simultaneously, these numbers are limiting values of the mean field
components $\mu_3$ and $\mu_8$ at zero temperature. For a degenerate
orbit one has to assign $m\,{=}\,0$, or $m\,{=}\,\sqrt{3}\, q$.
Evidently the degenerate orbits with $m\,{=}\,0$ are domains of mean
field configurations that realize nematic states. Ferromagnetic
states are realized on all other orbits of SU(3).

The group SU(4) has four types of orbits: the generic
$\frac{\mathrm{SU}(4)}{\mathrm{U}(1)\times \mathrm{U}(1)\times
\mathrm{U}(1)}$ of dimension $12$, the degenerate
$\frac{\mathrm{SU}(4)}{\mathrm{SU}(2)\times \mathrm{U}(1)\times
\mathrm{U}(1)}$ of dimension 10, the degenerate
$\frac{\mathrm{SU}(4)}{\mathrm{S}(\mathrm{U}(2)\times
\mathrm{U}(2))}$ of dimension 8, and the maximal degenerate
$\frac{\mathrm{SU}(4)}{\mathrm{SU}(3)\times \mathrm{U}(1)}$ of
dimension 6. Each orbit is defined by numbers $m$, $q$, $p$, which
are limiting values of the mean field components $\mu_3$, $\mu_8$,
$\mu_{15}$.  Almost all orbits are domains of ferromagnetic mean
field configurations. Nematic states are realized on the degenerate
orbits of dimension 8 as $m\,{=}\,p\,{=}\,0$ and $q$ is arbitrary.
So it is probable to reveal a nematic state even in the
ferromagnetic region of the phase diagram~\cite{Buchta}.

As shown above, \emph{limiting values of diagonal components of a
mean field} serve as order parameters. Simultaneously, they
\emph{define a coadjoint orbit} where the corresponding mean filed
configuration lives.

Each orbit possesses a Hamiltonian system with an equation of motion
and a group-invariant Hamiltonian. For a degenerate orbit of SU(3)
the equation is
\begin{equation}\label{HamEq}
\frac{\partial \mu_a}{\partial t} =
\tfrac{4\mathcal{A}}{3(m^2+q^2)}\, C_{abc} \mu_b ( \mu_{c,xx} +
\mu_{c,yy}),
\end{equation}
where $\mathcal{A}$ denotes a dimensional constant. The values $m$,
$q$ of order parameters (a magnetization and a projection of
quadrupole moment) define an orbit via the following equations,
which we call constraints:
$$\begin{array}{l}
\delta_{ab} \mu_a \mu_b = m^2+q^2\\ d_{abc} \mu_b \mu_c = \pm
\sqrt{\frac{m^2+q^2}{5}}\, \mu_a.
\end{array}$$ Here $d_{abc}\,{=}\, \frac{\sqrt{3}}{4\sqrt{5}}\Tr (\hat{P}_a\hat{P}_b\hat{P}_c+
\hat{P}_b\hat{P}_a\hat{P}_c)$ is a symmetric tensor. The
corresponding SU(3)-invariant Hamiltonian is the following
\begin{equation}\label{HamEff1}
\mathcal{H}_{\text{eff}}^{\text{deg}} =
\tfrac{2\mathcal{J}}{3(m^2+q^2)} \int \sum_{a=1}^8
\Bigl((\mu_{a,x})^2
  +(\mu_{a,y})^2\Bigr)\, dxdy,
\end{equation}
where the dimensional constant $\mathcal{J}$ has a meaning of
exchange integral.

Obviously, equations \eqref{HamEq} and \eqref{LLEq} coincide. It is
easy to show, that \eqref{LLEq} coincides with the equation of
motion on a maximal degenerate orbit. Recall, that \eqref{LLEq} is
obtained when correlations between fluctuations of $\{\hat{P}^a_n\}$
are neglected. Presumably, equations of motion on other orbits are
derived from \eqref{HeisEq} via the mean field averaging with more
complicate correlation rules.

A generic orbit of SU(3) is determined by the following equations:
$$\begin{array}{l}
\delta_{ab} \mu_a \mu_b = m^2+q^2\\ d_{abc}\mu_a \mu_b \mu_c =
\frac{1}{\sqrt{5}}\,q(3m^2-q^2).
\end{array}$$ The SU(3)-invariant Hamiltonian on this orbit
has the form
\begin{multline}\label{HamEff2}
\mathcal{H}_{\text{eff}}^{\text{gen}} = \tfrac{\mathcal{J}}{2
m^2(m^2-3q^2)^2} \sum_{a=1}^8 \Bigl( (m^2+q^2)^2 (\mu_{a,x})^2
  +(m^2+q^2)(\eta_{a,x})^2 - \\ - 2\sqrt{3}\,q(3m^2-q^2)  \mu_{a,x} \eta_{a,x}
  \Bigr),
\end{multline}where $\eta_{a}$ is a quadratic form in $\{\mu_a\}$:
$\eta_a \,{=}\, \sqrt{5}\, d_{abc}\mu_b \mu_c$. For more details
see~\cite{BernHolod}.

The Hamiltonian systems on coadjoint orbits of SU(3) serve as
\emph{classical effective models} for the spin system of
$s\,{\geqslant 1}\,$ with biquadratic exchange. Evidently, these
models describe large-scale (or slow) fluctuations of the mean
field. In this paper we suppose that the order parameters $m$, $q$
are fixed numbers. But generally speaking, they depend on a
temperature~$T$ and the interaction constant~$J$.  Taking into
account these dependencies, one can consider small-scale (or quick)
fluctuations of the mean field.

\subsection{Geometrical properties of effective Hamitonians} Each
coadjoint orbit of a semisimple Lie group is a homogeneous space
that admits a K\"{a}hlerian structure. Thus one can introduce a
complex parameterization of an orbit. For this purpose we use a
\emph{generalized stereographic projection} (for more details see
\cite{Bernatska}).
 In the case of group
SU(3), the projection is represented by the following formulas:
$$\mu_a = -\tfrac{m-\sqrt{3}\,q}{2}\, \zeta_a + m\xi_a,\quad \eta_a
= \tfrac{\sqrt{3}\,(m^2-q^2)-2mq}{2}\, \zeta_a + 2mq\xi_a,$$

\vspace{-10pt} \scriptsize
\begin{align}\label{StereoProj}
&\zeta_1 =
-\frac{w_2+\bar{w}_2+w_3+\bar{w}_3}{\sqrt{2}(1+|w_2|^2+|w_3|^2)}&
&\xi_1= -\frac{(1-\bar{w}_1)(w_3-w_1w_2)+(1-w_1)(\bar{w}_3-\bar{w}_1\bar{w}_2)}
{\sqrt{2}(1+|w_1|^2+|w_3-w_1w_2|^2))}&\nonumber\\
&\zeta_2= i
\frac{w_3-\bar{w}_3-w_2+\bar{w}_2}{\sqrt{2}(1+|w_2|^2+|w_3|^2)}&
&\xi_2 = i
\frac{(1+\bar{w}_1)(w_3-w_1w_2)-(1+w_1)(\bar{w}_3-\bar{w}_1\bar{w}_2)}
{\sqrt{2}(1+|w_1|^2+|w_3-w_1w_2|^2)}&\nonumber\\
&\zeta_3 = \frac{|w_2|^2-|w_3|^2}{1+|w_2|^2+|w_3|^2}& &\xi_3=
\frac{1-|w_1|^2}{1+|w_1|^2+|w_3-w_1w_2|^2}&\nonumber\\
&\zeta_4 = i \frac{\bar{w}_2w_3-w_2\bar{w}_3}{1+|w_2|^2+|w_3|^2}&
&\xi_4 = i\frac{w_1-\bar{w}_1}{1+|w_1|^2+|w_3-w_1w_2|^2}&\\
&\zeta_5 = \frac{w_2+\bar{w}_2
-w_3-\bar{w}_3}{\sqrt{2}(1+|w_2|^2+|w_3|^2)}&
&\xi_5=-\frac{(1+\bar{w}_1)(w_3-w_1w_2)+(1+w_1)(\bar{w}_3-\bar{w}_1\bar{w}_2)}
{\sqrt{2}(1+|w_1|^2+|w_3-w_1w_2|^2)}
&\nonumber\\
&\zeta_6 = i
\frac{w_2-\bar{w}_2+w_3-\bar{w}_3}{\sqrt{2}(1+|w_2|^2+|w_3|^2)}&
&\xi_6 =
i\frac{(1-\bar{w}_1)(w_3-w_1w_2)-(1-w_1)(\bar{w}_3-\bar{w}_1\bar{w}_2)}
{\sqrt{2}(1+|w_1|^2+|w_3-w_1w_2|^2)}\nonumber\\
&\zeta_7= - \frac{\bar{w}_2w_3+w_2\bar{w}_3}{1+|w_2|^2+|w_3|^2}&
&\xi_7=-\frac{w_1+\bar{w}_1}{1+|w_1|^2+|w_3-w_1w_2|^2}\nonumber\\
&\zeta_8 =\frac{2-|w_2|^2-|w_3|^2}{\sqrt{3}(1+|w_2|^2+|w_3|^2)}&
&\xi_8= \frac{1+|w_1|^2-2|w_3-w_1w_2|^2}
{\sqrt{3}(1+|w_1|^2+|w_3-w_1w_2|^2)}.& \nonumber
\end{align}
\normalsize The coordinates $\{w_1,\, w_2,\, w_3\}$ (Bruhat
coordinates according to \cite{Picken}) parameterize a generic orbit
of $\mathrm{SU}(3)$. In the case of a degenerate orbit, one has to
assign $m\,{=}\,0$ and $w_1\,{=}\,0$, or $m \,{=}\, \sqrt{3}\,q$ and
$w_2\,{=}\,0$.

In terms of $\{w_\alpha\}$ the effective Hamiltonians
\eqref{HamEff1} and \eqref{HamEff2} have the form
\begin{align}\label{MetricsHam}
  &\mathcal{H}_{\text{eff}} = \mathcal{J} \int \sum_{\alpha,\beta} g_{\alpha\bar{\beta}} \Bigl(
  \frac{\partial w_{\alpha}}{\partial z}\frac{\partial \bar{w}_{\beta}}{\partial \bar{z}}+
  \frac{\partial w_{\alpha}}{\partial \bar{z}}
  \frac{\partial \bar{w}_{\beta}}{\partial z}\Bigr)\, dz
  d\bar{z},\quad \alpha,\, \beta=1,\,2,\,3,
\end{align}
where $z=x+iy$ is a complex coordinate on the plane obtained from
the atomic lattice after a large-scale limiting process (see
Section~\ref{s:MFA}). The tensor $g$ is non-degenerate and
positively defined, thus it can serve as a metrics on an orbit. Its
components $\{g_{\alpha\bar{\beta}}\}$ come from \eqref{HamEff1} for
a degenerate orbit, and from \eqref{HamEff2} for a generic one. In
terms of the auxiliary vector fields $\{\zeta_a\}$ and $\{\xi_a\}$
we have
\begin{align*}
&g^{\rm deg 1}_{\alpha \bar{\beta}}= \tfrac{1}{2}\sum_{a}
\frac{\partial \zeta_a}{\partial w_{\alpha}}\frac{\partial
\zeta_a}{\partial \bar{w}_{\beta}},\qquad g^{\rm deg 2}_{\alpha
\bar{\beta}}=  \tfrac{1}{2}\sum_{a} \left.\frac{\partial
\xi_a}{\partial w_{\alpha}}\frac{\partial
\xi_a}{\partial\bar{w}_{\beta}}\right|_{w_2=0},\\
&g^{\rm gen}_{\alpha \bar{\beta}}= \tfrac{1}{2}\sum_{a} \left(
\frac{\partial \zeta_a}{\partial w_{\alpha}}\frac{\partial
\zeta_a}{\partial \bar{w}_{\beta}} - \frac{\partial
\zeta_a}{\partial w_\alpha}\frac{\partial \xi_a}{\partial
\bar{w}_\beta} + \frac{\partial \xi_a}{\partial
w_\alpha}\frac{\partial \xi_a}{\partial \bar{w}_\beta} \right).
\end{align*}
Note, that in terms of $\{w_\alpha\}$ the tensor $g$ does not depend
on a particular orbit.

Being a K\"{a}hlerian manifold an orbit of SU(3) possesses a
K\"{a}hlerian potential. For this purpose we use a potential $\Phi$
of the Kirillov-Kostant-Suoriau form:
\begin{gather*}
\Phi = m\Phi_1 - \tfrac{m-\sqrt{3}\,q}{2}\,\Phi_2,\\ \Phi_1 =
\ln(1+|w_1|^2 + |w_3{-}w_1 w_2|^2),\quad \Phi_2 =
\ln(1+|w_2|^2+|w_3|^2),
\end{gather*}
A topological structure of the orbit is characterized by the second
cohomology group $H^2$ of dimension 2. That is why there exist two
basis 2-forms, for example generated by the potentials $\Phi_1$,
$\Phi_2$. Each of them defines a topological charge
\begin{equation*}
  \mathcal{Q}_k = \frac{1}{4\pi}\int \sum_{\alpha,\beta}
  \frac{i\partial^2 \Phi_k} {\partial w_{\alpha} \partial \bar{w}_{\beta}} \Bigl(
  \frac{\partial w_{\alpha}}{\partial z}\frac{\partial \bar{w}_{\beta}}{\partial
  \bar{z}}-
  \frac{\partial w_{\alpha}}{\partial \bar{z}}\frac{\partial \bar{w}_{\beta}}{\partial z}\Bigr)\,
  dz{\wedge} d\bar{z},\quad k=1,\,2.
\end{equation*}

On a degenerate orbit only one potential is governing, and only one
to\-po\-lo\-gi\-cal charge exists. Then the expressions for
$\mathcal{Q}_k$ and $\mathcal{H}_{\text{eff}}^{\text{deg}\,k}$
differ only in a sign. Evidently,
\begin{equation}\label{MinRel}
\mathcal{H}_{\text{eff}}^{\text{deg}\,k} \geqslant 4\pi \mathcal{J}
|\mathcal{Q}_k|.
\end{equation}
Hence, \emph{on a degenerate orbit a minimum of
$\mathcal{H}_{\text{eff}}$ is realized if the equality holds, that
takes place  if $\{w_{\alpha}\}$ are holomorphic or antiholomorphic
functions}. Here we use an idea of Belavin and Polyakov
\cite{Belavin}.

For a generic orbit we define a topological charge by $\mathcal{Q}
\,{=}\, \mathcal{Q}_1\,{+}\,\mathcal{Q}_2$. In order to extend
inequality \eqref{MinRel} to generic orbits with the topological
charge $\mathcal{Q}$, we construct an effective Hamiltonian by the
following formula:
\begin{gather}\label{HamEff2D}
\mathcal{H}_{\text{eff}}^{\text{gen}} = \tfrac{\mathcal{J}}{2
m^2(m^2-3q^2)^2} \sum_{a=1}^8 \Bigl( C_1 (\mu_{a,x})^2
  + C_2(\eta_{a,x})^2 + C_3 \mu_{a,x} \eta_{a,x}
  \Bigr), \\ \begin{array}{l}
  C_1 = m^4+q^4-\frac{4}{\sqrt{3}}\,mq(q^2-m^2)+14m^2q^2,\\
  C_2 = \frac{5}{3}\,m^2+q^2 - \frac{2}{\sqrt{3}}\,mq,\\
  C_3 = \frac{2}{\sqrt{3}}\, m^3 + 2q^3 -\frac{26}{3}\,m^2q +
  \frac{2}{\sqrt{3}}\,mq^2.
  \end{array} \nonumber
\end{gather}
In terms of $\{w_\alpha\}$ it is reduced to the form
\eqref{MetricsHam} with the  metrics
$$g^{\rm gen}_{\alpha \bar{\beta}}= \tfrac{1}{2}\sum_{a} \left(
\frac{\partial \zeta_a}{\partial w_{\alpha}}\frac{\partial
\zeta_a}{\partial \bar{w}_{\beta}} + \frac{\partial \xi_a}{\partial
w_\alpha}\frac{\partial \xi_a}{\partial \bar{w}_\beta} \right).$$
Then we get $$ \mathcal{H}_{\text{eff}}^{\text{gen}} \geqslant 4\pi
\mathcal{J} |\mathcal{Q}|,$$ and \emph{a minimum of
$\mathcal{H}_{\text{eff}}^{\text{gen}}$ is realized if the equality
holds, that takes place  if $\{w_{\alpha}\}$ are holomorphic or
antiholomorphic functions}.

%------------------------------------------------------------------------------
\section{Large-scale topological excitations}
Now we construct a particular class of topological excitations that
give minimums to the effective Hamiltonians
$\mathcal{H}_{\text{eff}}^{\text{deg}\,1}$, and
$\mathcal{H}_{\text{eff}}^{\text{gen}}$ defined by \eqref{HamEff2D}.
We describe these excitations by holomorphic functions
$\{w_{\alpha}(z)\}$. Each set $\{w_{1}(z),$ $w_{2}(z)$, $w_{3}(z)\}$
represents a mean field configuration of the system in question.

First, we consider a degenerate orbit of~$\mathrm{SU}(3)$ with
$m\,{=}\,0$, $q\,{=}\,{-}\frac{2}{\sqrt{3}}$, where a nematic state
is realized. In order to satisfy the limiting conditions:
$\mu_3\,{\to}\, m$, $\mu_8\,{\to}\, q$, the functions
$\{w_{\alpha}(z)\}$ have to vanish as $z\,{\to}\, \infty$. Let
\begin{equation}\label{Excit1}
w_1(z)=0,\quad w_2(z) {=} \tfrac{a_2}{z-z_2},\quad w_3(z) {=}
\tfrac{a_3}{z-z_3},\qquad a_2,\, z_2,\,a_3,\,z_3\in \Complex,
\end{equation}
be a large-scale excitation in the magnet in question.  The
corresponding mean field configuration is obtained by substitution
of \eqref{Excit1}  into \eqref{StereoProj}. A behavior of the
components $\mu_3$ and $\mu_8$  is represented on the Fig.\,1. As
$z\,{\to}\, \infty$ the values of $\mu_3$, $\mu_8$ tend to $m$, $q$.

\begin{figure}[h]
\centering
\includegraphics[width=0.8\textwidth]{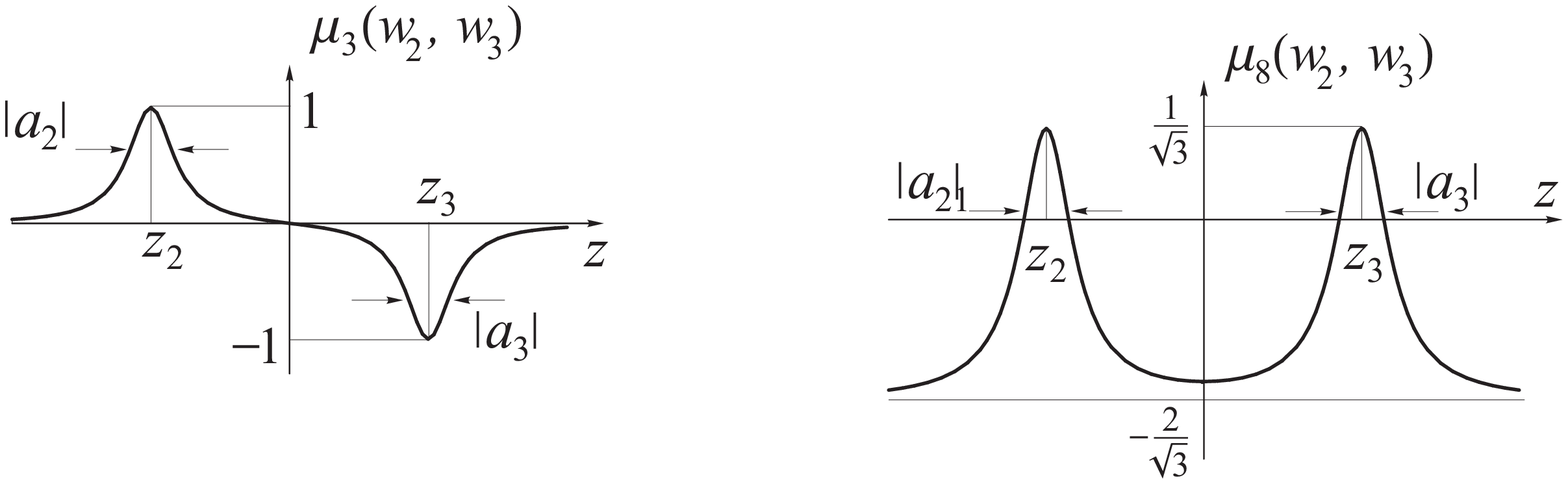}\\
Fig.~1. Profiles of the mean field components $\mu_3$, $\mu_8$ in
the case of configuration \eqref{Excit1}.
\end{figure}

Suppose $z_2=z_3$. By shifting of the coordinate $z$ one can easily
reduce \eqref{Excit1} to the configuration
\begin{equation*}
w_1(z)=0,\quad w_2(z) {=} \tfrac{a_2}{z},\quad w_3(z) {=}
\tfrac{a_3}{z},\qquad a_2,\, a_3\in \Complex,
\end{equation*}
and calculate its topological charge:
\begin{equation}\label{TopCh1}
\mathcal{Q} = \frac{2i}{4\pi}\iint_{\Complex} \frac{(a_2^2 +
a_3^2)}{(|z|^2+a_2^2 + a_3^2)^2}\; dz\wedge d\bar{z} = 1.
\end{equation}
In the case of $z_2 \neq z_3$ the topological charge equals 2.

It can be interpreted as follows. Each pole of a mean field
configuration represents a kind of Belavin-Plyakov soliton. Each
soliton gives a unit topological charge. Thus, two distinct solitons
have the topological charge 2. No continuous deformation take a
configuration of topological charge 2 to a configuration of
topological charge 1. If we allow noncontinuous deformation, then
two solitons can meet at any point and join into one, at the same
time an energy is released. From \eqref{MinRel}, it follows that the
released energy equals $4\pi\mathcal{J}$ per one pole.

Note, that the energy of configuration \eqref{Excit1} does not
depend on parameters of solitons: $a_2$, $z_2$, $a_3$, $z_3$. It
means that the excitation can grow (when $|a_2|$ and $|a_3|$ grow)
preserving an energy. Such growth immediately leads to destruction
of an order in the system.

One can construct a configuration with more than two solitons:
\begin{equation}\label{Excit2}
w_1(z)=0,\quad w_2(z) = a_2 / \prod_{k=1}^n (z-z_{2k}),\quad w_3(z)
= a_3 / \prod_{k=1}^m (z-z_{3k}),
\end{equation}
here $a_2$, $a_3$, $\{z_{2k}\}_{k=1}^n$, and $\{z_{3k}\}_{k=1}^m$
are fixed complex numbers. If all values $\{z_{2k}\}_{k=1}^n$ and
$\{z_{3k}\}_{k=1}^m$ are distinct, a topological charge equals
$n\,{+}\,m$. When a pole of the function $w_2(z)$ coincides with a
pole of $w_3(z)$, the topological charge decreases by 1. But a
coincidence of two poles of the same function (for example $w_2(z)$)
does not lead to a decrease of the topological charge. It is easy to
see, that the minimal energy of configuration \eqref{Excit2} equals
$4\pi \mathcal{J}\cdot\min(n,m)$.

Now we consider a generic orbit, where a ferromagnetic state is
realized. Suppose $m\,{=}\,1$ and $q\,{=}\,{-}\frac{2}{\sqrt{3}}$.
In this case, we describe a mean field in the magnet by the
effective Hamiltonian $\mathcal{H}_{\text{eff}}^{\text{gen}}$,
defined by \eqref{HamEff2D}. Let
\begin{equation}\label{Excit3}
w_1(z)=\tfrac{a_1}{z-z_1},\quad w_2(z) {=} \tfrac{a_2}{z-z_2},\quad
w_3(z) {=} \tfrac{a_3}{z-z_3},\quad a_k, z_k \in \Complex,\ \
k=1,2,3.
\end{equation}
be a large-scale excitation of the mean field. A calculation of
topological charges gives: 1) $\mathcal{Q}_1\,{=}\,3$,
$\mathcal{Q}_2\,{=}\,2$, if $a_1\,{\neq}\,a_2\,{\neq}\,a_3$ or
$a_1\,{=}\,a_2\,{\neq}\,a_3$, 2) $\mathcal{Q}_1\,{=}\,2$,
$\mathcal{Q}_2\,{=}\,2$, if $a_1\,{=}\,a_3\,{\neq}\,a_2$, 3)
$\mathcal{Q}_1\,{=}\,2$, $\mathcal{Q}_2\,{=}\,1$, if
$a_1\,{=}\,a_2\,{=}\,a_3$ or $a_1\,{\neq}\,a_2\,{=}\,a_3$.

\section{Conclusion and discussion}
Each generalized Heisenberg Hamiltonian with high powers of the
exchange interaction can be reduced to a bilinear form. By a mean
field averaging  we obtain a classical system from the original
quantum one. An ave\-ra\-ging of the Heisenberg equation gives a
Landau-Lifshitz like equation for a mean field. Using Lie group
apparatus, we construct effective Hamiltonians for the classical
system with SU(3) symmetry. One of them
$\mathcal{H}_{\text{eff}}^{\text{deg}}$ is an SU(3)-analogue of the
Hamiltonian commonly used in theory of magnetism. In addition, we
propose another one $\mathcal{H}_{\text{eff}}^{\text{gen}}$, which
is biquadratic in the mean field. Further, we construct examples of
topological excitations that give minimums to the Hamiltonians. Such
excitations can change their shapes and grow preserving an energy.
This is a probable scenario for the destruction of an ordered state
in a 2D magnet at nonzero temperature, that agrees with the
Mermin-Wagner theorem.

\section{Acknowledgments}
This work is partly supported by the grant of the International
Charitable Fund for Renaissance of Kyiv-Mohyla Academy.

\end{document}